\documentstyle[11pt,aaspp4]{article}
 
\begin{document}

\title{Re-examining the Lyman Continuum in Starburst Galaxies Observed with
the Hopkins Ultraviolet Telescope}

\author{Mark Hurwitz, Patrick Jelinsky, and W. Van Dyke Dixon}
\affil{Space Sciences Laboratory,
University of California,\\ Berkeley, CA 94720-7450}
%\altaffiltext{1}{also with Space Sciences Laboratory,
%University of California,\\ Berkeley, CA 94720-7450}

\authoremail{markh@ssl.berkeley.edu}

\begin{abstract}
We have reevaluated the constraints on the Lyman continuum
emission from four starburst galaxies observed with
the Hopkins Ultraviolet Telescope (HUT) during the Astro-2
mission.  Applying a detailed model of the absorption 
by interstellar gases in our Galaxy, and using
the latest HUT data products, we find upper limits to
the redshifted Lyman continuum that are less restrictive than 
those reported previously (Leitherer et al.\  1995 ApJ, 454, L19).
Well determined astrophysical and 
instrumental effects permit 2 $\sigma$ upper limits 
no tighter than 5.2\%, 11\%, 57\%, and 3.2\% to be set
on the escape fraction of Lyman continuum
photons from IRAS 08339+6517, Mrk 1267, Mrk 66, and Mrk 496, respectively.
Absorption from undetected interstellar components
(including H$_2$) or modulation of the emergent
spectrum by gas or dust in the parent galaxy
could allow the true escape fractions 
to exceed these revised upper limits.

\end{abstract}

%\keywords{Galaxies: Starburst}

\section{INTRODUCTION}

In a recent study, Leitherer et al.\  (1995) set
upper limits on the fraction of Lyman continuum photons
escaping from four starburst galaxies observed with
the Hopkins Ultraviolet Telescope (HUT).
The 2~$\sigma$ upper limit, expressed as a fraction of
the number of ionizing photons produced (estimated
from Balmer emission line strengths), was typically 
found to be about 3\%.  
Star-forming regions in galaxies of this type
are generally heavily reddened; the escape of
Lyman continuum photons thus depends sensitively on the 
patchiness of the absorbers in the star-forming galaxy.
Because the patchiness is difficult to constrain
independently, the study of Leitherer et al.\ 
is important despite the small sample size.

The emergent Lyman continuum flux can be constrained
experimentally by setting limits on the observed 
source brightness between 915 \AA\ and the redshifted
Lyman limit of the galaxy under study
(which falls between 929 and 939 \AA\ for this sample).
Leitherer et al.\  assume that the intrinsic spectrum is 
flat in this region and that Galactic absorption, other than 
extinction by dust, can be neglected.  In this work we address
the last assumption in some detail and demonstrate
that Galactic gas-phase absorption attenuates a significant
fraction of the flux between 915 \AA\ and the redshifted
Lyman limit along these sight lines, leading to
looser constraints on the fraction of
ionizing photons that may escape the starburst galaxy.

\section{MODELING THE GALACTIC ABSORPTION}

The basis for our absorption models is the Bell Labs 21 cm survey 
(Stark et al.\  1992), which maps $N_{\rm H\,I}$,
the primary gas phase absorber at the wavelengths of interest,
in 10 km s$^{-1}$ velocity channels.
For each sight line we adopt the 21 cm profile from the
nearest beam of the survey; these profiles are
shown in Figure 1.
To approximate the smooth distribution of H~I with velocity,
we assign a Doppler $b$ value of 4.3 km s$^{-1}$ 
to a numerical ``cloud'' centered at each velocity channel.
We adopt a profile for O~I identical to that of H~I,
but with a column that is lower by 3.3 dex (de Boer 1981).

In physical clouds, the $b$ values of the heavier 
elements may well be lower than that of H~I\@.
However, the broad range of velocities at 
which H~I is observed must be dominated by bulk motions or
turbulence rather than thermal widths.  The O~I contributes
only about 7\% of the total equivalent width of the gas phase
absorption, so even fairly gross errors in its velocity 
distribution would have a small effect on our final results. 
Other atomic species with absorption features in 
the bandpass of interest include N~II, N*~II, Ar~II, Fe~II, Al~II, and S~VI;
we neglect these on the grounds that their contribution
to the absorption will be fairly low and/or that their 
column is difficult to predict.

To calculate the far ultraviolet absorption 
expected from a given 21 cm profile,
we establish a very fine wavelength grid (0.001 \AA) 
and calculate exact Voigt profiles for the species and velocity
components along the sight line.  We co-add the
optical depths, exponentiate, and
convolve with a 3~\AA\ FWHM Gaussian to approximate
the HUT resolution.
(Small differences between our assumed resolution
and the actual value will have no effect on the end result.)
We then bin the transmission
on a 0.51 \AA\ grid to match the HUT detector pixels.
We show a sample interstellar absorption profile,
calculated for IRAS 08339+6517, in Figure 2.

To assess the statistical uncertainty in the far
ultraviolet gas phase absorption, we vary the 
H I column in each velocity channel in Monte Carlo
fashion, assuming a Gaussian distribution 
centered on the measured value but with
a 1~$\sigma$ uncertainty of $10^{17}$ cm$^{-2}$
in each channel.  This uncertainty is typical of the empirical
excursions in the wings of the 21 cm line,
which extend out to $\pm$ 360 km s$^{-1}$.
The true probability of encountering a $10^{17}$ cm$^{-2}$ cloud 
at a given velocity is not easy to predict,
but, at least at the velocity extrema, it is
presumably lower than would be estimated from the measurement error.
We therefore crop the profile beyond 30 km s$^{-1}$ from the 
velocity range within which H~I is clearly detected.
The resulting interstellar transmissions and
statistical uncertainties are shown in Table 1.

\begin{deluxetable}{llllllrllllll}
\footnotesize
\tablewidth{468pt}
\tablecaption{Summary of analysis of the four starburst
galaxies observed with HUT.}
\tablehead{
\colhead{Galaxy}&
\colhead{Integ\tablenotemark{a}} &
\colhead{Src\tablenotemark{b}}& 
\colhead{Err\tablenotemark{c}}&
\colhead{BG\tablenotemark{d}}&
\colhead{Err\tablenotemark{e}}& 
\colhead{Net\tablenotemark{f}} & 
\colhead{Err\tablenotemark{f}} &
\colhead{Band\tablenotemark{b}} &
\colhead{$T_{\rm ism}$\tablenotemark{g}} & 
\colhead{Err} &
\colhead{Flux\tablenotemark{h}}&
\colhead{$F_{\rm esc}$\tablenotemark{i}} \nl \cline{3-8}
& \colhead{(s)}&\multicolumn{6}{c}{(counts)}& \colhead{(\AA)}&&&
& \colhead{(\%)}\nl
 \colhead{(1)} &\colhead{(2)} & \colhead{(3)} & \colhead{(4)} & \colhead{(5)} & 
\colhead{(6)} & \colhead{(7)} & \colhead{(8)} & \colhead{(9)} & \colhead{(10)} & 
\colhead{(11)} & \colhead{(12)}& \colhead{(13)}}
\startdata
IRAS 08339 &2752 & 41 & 6.4 &43.3 &3.4 &$-2.3$& 7.3 &13.8 &0.44 & 0.014
&$<1.7$ &$<\phn4.1$ \nl
\phm{IRAS}+6517  \nl
Mrk 1267   &3062 & 37 & 6.1 &34.9 &3.2 & $2.1$& 6.9 &14.3 &0.56 & 0.013
&$<1.4$ &$<\phn8.5$ \nl
Mrk 66     &2176 & 40 & 6.3 &41.4 &3.7 &$-1.4$& 7.3 &16.8 &0.49 & 0.017
&$<1.6$ &$<43 $ \nl
Mrk 496    &1410 & 51 & 7.1 &52.1 &4.9 &$-1.1$& 8.7 &23.5 &0.63 & 0.014
&$<1.6$ &$<\phn2.3$ \nl
\tablenotetext{a}{Effective integration time in seconds.}
\tablenotetext{b}{Raw detector counts from 915 \AA\ to the redshifted Lyman limit;
the width of the band is shown in column 9.}
\tablenotetext{c}{Square root of column 3.}
\tablenotetext{d}{Background expected in this wavelength interval, scaled
from raw counts in the $850 - 900$ \AA\ region.}
\tablenotetext{e}{Statistical error in column 5.}
\tablenotetext{f}{Net counts and statistical error therein.}
\tablenotetext{g}{Mean interstellar transmission within the band, including only the
known atomic species.}
\tablenotetext{h}{The 2~$\sigma$ maximum
permitted flux at Earth, corrected for gas phase absorption
(but not for extinction, for ease of comparison with Leitherer et al.\  1995).
Units are $\times10^{-15}$ erg (cm$^2$ s \AA)$^{-1}$.}
\tablenotetext{i}{Corresponding limit on the escape fraction
for ionizing photons from the galaxy, scaled from Leitherer et al.
Note that the limits in columns 12 and 13 may be
unrealistically low for reasons discussed in the text.}
\enddata
\end{deluxetable}

\section{DATA REDUCTION AND ANALYSIS}

The observations of starburst
galaxies were performed 
with HUT on the Astro-2 mission of the
space shuttle {\it Endeavour} in 1995 March. HUT is a far-UV
spectrophotometer with first-order sensitivity extending from 820 to
1840 \AA\ at 0.51 \AA\ pixel$^{-1}$; its resolution ranges
from 2.1 to 2.5  \AA\ FWHM between 900 and 950 \AA. 
The spectrograph and telescope are described in detail by
Davidsen et al.\  (1992), while Kruk et al.\  (1995) discuss its
performance and calibration on the Astro-2 mission.

Four starburst galaxies, IRAS 08339+6517, Mrk~1267, Mrk~66, and Mrk~496
(= NGC 6090), were observed several times during the Astro-2 mission.
Basic parameters of the galaxies and their flux-calibrated HUT spectra
may be found in Leitherer et al.\  (1995).  For each observation, data
products from the HUT ``Ballistic Process'' data-reduction system have
been kindly provided by G. Kriss. We exclude from our analysis
observations in which the source appears to spend 
more than 50\% of the observation outside of the aperture, based on
plots of source count rate vs. time. 
All the observations included here were taken
through the 20\arcsec\ circular aperture.
Spectra from multiple observations were co-added; total 
integration times are listed in
Table~1.  For Mrk 496, we have scaled the integration
time by 0.8 to account for excursions of the target 
from the aperture, bringing the HUT flux at longer wavelengths
into agreement with the {\it IUE} measurements
of Kinney et al.\  (1993).

We carry out most of our analysis of the spectra using raw 
counts rather than calibrated flux units, because
detector dark counts and stray light (which
dominate the signal below the redshifted Lyman limit) 
are expected to be ``flatter'' in raw counts.
We sum the source counts from 915.0 \AA\ to
the redshifted Lyman limit, adopting the radial velocities
from Leitherer et al.  We estimate the background in
this interval by scaling the summed counts from 850
to 900 \AA.

Treating both the background-subtracted source counts and
the mean interstellar transmission as independent
Gaussian variables with $\sigma$ shown in Table 1,
we can arrive at a 95\% upper limit on the photon
flux incident on the Galaxy (but not yet corrected 
for Galactic dust extinction), which we then multiply by the photon energy,
and divide by the product of the integration time, effective area,
and bandpass in \AA.
The results are summarized in Table 1, where the
final column (escape fraction) is scaled from
Leitherer et al., assuming the Galactic extinction
and H$\alpha$ luminosity for each galaxy from the previous work.
 
To explore the sensitivity of our results to
the detailed choice of statistical technique, we
have applied two alternative methods.  The first
is an analysis similar to the one described above,
except we first exclude those wavelength bins
for which the mean gas-phase interstellar transmission 
is less than 0.5 (from 915 to, typically, 920 \AA).
This reduces the effective
bandpass of the exposure, but raises the 
mean interstellar transmission over the remaining
interval.  In all cases, the final result 
is within 12\% of the value cited in Table 1.
We have also applied a maximum likelihood method.
Here the alternative statistical treatment
sometimes yields a more significant difference; 
the largest is for Mrk 1267, where the flux 
constraint is 75\% of the value in Table 1.

\section{OTHER SOURCES OF UNCERTAINTY}

We now consider whether effects other than 
the shot noise in the detector counts could 
affect our limits on the Lyman continuum escape fraction.

Our estimate of the per-pixel
background, comprising detector dark counts and stray light,
is insensitive to the detailed endpoints
of the adopted extreme ultraviolet wavelength interval.
The background in the 840 -- 885 \AA\
band is consistent, within fairly small statistical
uncertainties, with that in the 850 -- 900 \AA\ band,
indicating that our background estimate is
not affected by bright airglow features near the endpoints.

To explore the potential variation of the background 
with wavelength, we analyzed ten HUT spectra for which the 
source was either negligibly faint or outside the aperture completely.
In four of four blank sky spectra collected through 
the $19\arcsec \times 197\arcsec$ aperture,
and in five of six spectra
collected through the the $10\arcsec \times 56\arcsec$ aperture,
the background is higher at 920 \AA\ than in the extreme ultraviolet.
The relative contributions of stray light and
detector dark count rate can be estimated from the
variation in the total background flux with aperture area.
Extrapolating to the 20\arcsec\ diameter aperture 
used in this work, we find that the detector dark counts,
not the stray light, contribute the vast majority (circa 80\%)
of the background near 920 \AA\ in these spectra.

The reliability of the background subtraction therefore 
depends primarily on the flatness of the detector dark counts spectrum.
Greeley (1995, private communication) has compiled 
nearly 11 on-orbit hours of dark counts data, which we
have searched for nonuniformities.
At phosphor voltage level 3, the dark spectrum
between 915 and 930 \AA\ is marginally (2\%) higher 
than the level between 850 and 900 \AA\, with a 1 $\sigma$ 
uncertainty corresponding to 8\% of the mean level.
At phosphor voltage level 4, there is evidence
for a genuine variation.  At this voltage the 
count dark spectrum between 915 and 930 \AA\ is 13\% lower
than the level between 850 and 900 \AA\, with a 1 $\sigma$ 
uncertainty corresponding to 5.5\% of the mean level.
More than half of the observations of IRAS 08339+6517,
and all of the observations of the remaining three
starbursts, were collected at phosphor voltage level 4.
Although this effect would not be credible in the absence
of independent data, columns 7 and 8 of Table 1 
provide some evidence for an oversubtraction
of the background flux.

We reanalyze the starburst spectra, correcting for
a 13\% of background systematic subtraction offset
and a 5.5\% of background statistical uncertainty
(added in quadrature with the shot noise).
Inclusion of these effects raises the 95\% upper limit 
on the escape fractions to 5.2\%, 11\%, 57\%, and 3.2\%
for IRAS 08339+6517, Mrk~1267, Mrk~66, and Mrk~496, respectively.

A second effect is the potential for additional 
gas phase absorption from the interstellar medium.
Our assessment of the statistical uncertainties in
the absorption is an oversimplified
treatment, given the complexities of the 21 cm baseline
subtraction, calibration, etc.
{\it The radio surveys cannot rule out the existence
of clouds whose convergent Lyman series would add
significant attenuation to the absorption curves
found here.}  Independent constraints on such
clouds would require high resolution, 
high signal-to-noise far-ultraviolet spectra, 
which will be challenging given the faintness of the sources.
For two of the galaxies (IRAS 08339 and Mrk 1267),
the known H~I column is sufficiently high that significant
H$_2$ is likely to be present (Savage et al.\  1977),
and for the other two galaxies, its presence cannot be ruled out.
The quantitative effect of additional H~I or H$_2$
clouds on the transmission of the interstellar gas depends 
on many parameters, all of them difficult to 
constrain independently. 
The authors' subjective assessment is that it would be
prudent to revise the previously listed
upper limits upward by a factor of about 1.25 
to account for additional interstellar components.
Even more substantial revisions may be necessary 
if the clouds are present over a wide velocity range.

Finally, we must remain cognizant of the potential for
interstellar absorption in the {\it parent} 
galaxies.  Leitherer et al.\  demonstrate that the production of
ionizing photons immediately below the Lyman limit
is well correlated with the production of ionizing photons
of all wavelengths for a broad range of stellar mass distributions.
Photoelectric absorption by H~I in the parent galaxy
will of course attenuate the longest wavelengths most strongly.
Dust grains can produce a similar effect, serving as
the primary cause of extinction immediately below the Lyman limit
if the escape paths are filled with highly
ionized hydrogen (e.g., less than about $4\times 10^{-4}$ neutral)
with a standard gas-to-dust ratio.  
Below the Lyman limit, interstellar grains tend
to absorb rather than scatter.  As the photon wavelength
decreases from 912 \AA\ to about 90 \AA,
the dust cross section falls by a factor
of about 3 (Martin \& Rouleau 1991).
Like H~I, dust preferentially absorbs 
photons immediately below the Lyman limit,
opening the possibility that the escape fraction
for higher energy photons could be significantly 
greater than the values derived from observations
of galaxies at these comparatively low velocities.

\section{DISCUSSION AND CONCLUSIONS}

We have shown that astrophysical and
instrumental effects permit 2 $\sigma$ upper limits
no tighter than 5.2\%, 11\%, 57\%, and 3.2\% to be set
on the escape fraction of Lyman continuum
photons from IRAS 08339+6517, Mrk 1267, Mrk 66, and Mrk 496, respectively.
Additional interstellar absorption would permit the 
true escape fraction to be higher than these limits.
The effect of such absorption is difficult to
quantify but a subjective assessment indicates
that a multiplicative factor of 1.25 would be prudent.
Differential absorption across the Lyman continuum 
by interstellar material in the parent galaxy would permit 
the true escape fraction to be higher still.

Madau \& Shull (1996) estimate, based on the 
metal enrichment of Ly$\alpha$ clouds,
that massive stars could contribute significantly
to the UV background at early epochs if
about 25\% or more of the Lyman continuum
photons can escape the star-forming regions.
Our revised upper limits are significantly closer 
to this escape fraction than were those reported
previously.  We have, thus, measurably reduced the 
weight of evidence---albeit preliminarily---against the 
hypothesis that Lyman continuum photons 
from massive stars contribute to the ionization 
of the universe at early epochs.

The complex absorption caused by the gas phase
of the ISM near the Galactic Lyman limit, and
the difficult-to-characterize uncertainties therein,
can be largely sidestepped by observing
starburst galaxies at higher velocity.
To be conservative, the region shortward
of about 930--940 \AA\ in the observer's
frame (depending on resolution) 
should be considered suspect.  Target galaxies with
redshifts greater than about 12,600 km s$^{-1}$ would
be highly desirable, although the potential for interstellar
absorption in the parent galaxy will remain a difficulty
even at these velocities.

\acknowledgements
We would like to thank G. Kriss for providing output products from the
HUT Ballistic Process data-reduction system. The Hopkins Ultraviolet
Telescope Project is supported by NASA contract NAS5-27000 to 
Johns Hopkins University.  We acknowledge the support
of NASA grant NAG5-696, and the helpful comments of
the anonymous referee.

\clearpage

% FIG 1
% \figcaption{H\,{\sc i} 21 cm profiles from Stark et al.\
% for the four starburst sight lines of $(a)$ IRAS 08339+6517, $(b)$ Mrk 1267,
% $(c)$ Mrk 66, and $(d)$ Mrk 496.
% Vertical dashed lines indicate the velocity range
% within which we assume that the H\,{\sc i} detections are secure.}

% FIG 2
% \figcaption{Gas phase interstellar medium transmission
% curve calculated for the sight line toward IRAS 08339+6517.
% Theoretical transmission has been convolved with a 3~\AA\
% FWHM Gaussian and rebinned at 0.51 \AA.
% Only H~I and associated metal lines are included.}

%%%%%%%%%%%%%%%%%%%%%%%%%%%%%%%%%%%%%%%%%%%%%%%%%%%%%%%%%%%%%%%%%%
% TO INCOPORATE FIGURES INTO PAPER, UNCOMMENT THE LINES BELOW
% AND COMMENT OUT THE \figcaption LINES ABOVE.  aef 2/13/97
%
% FIG 1
 \begin{figure}
 \plottwo{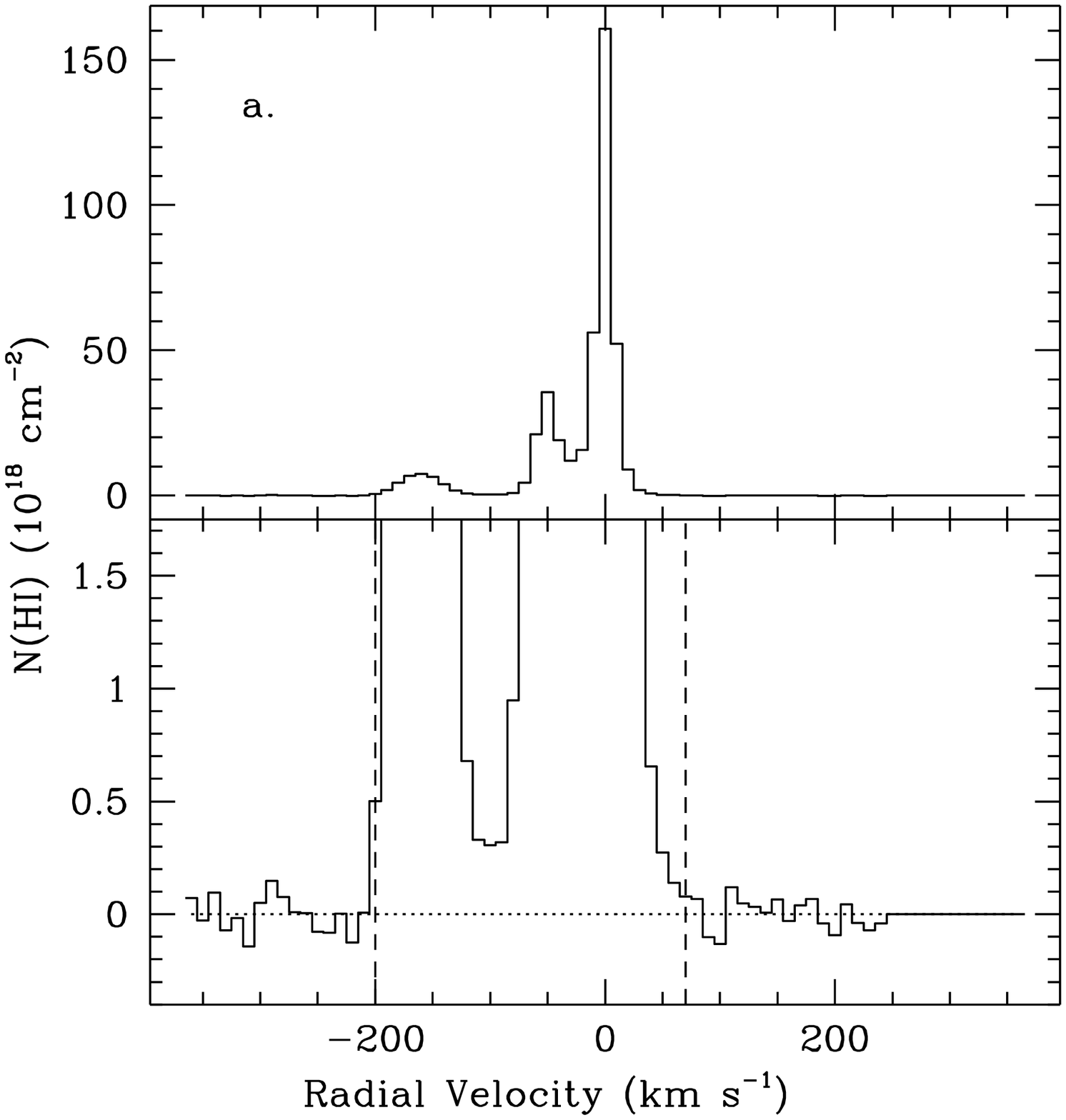}{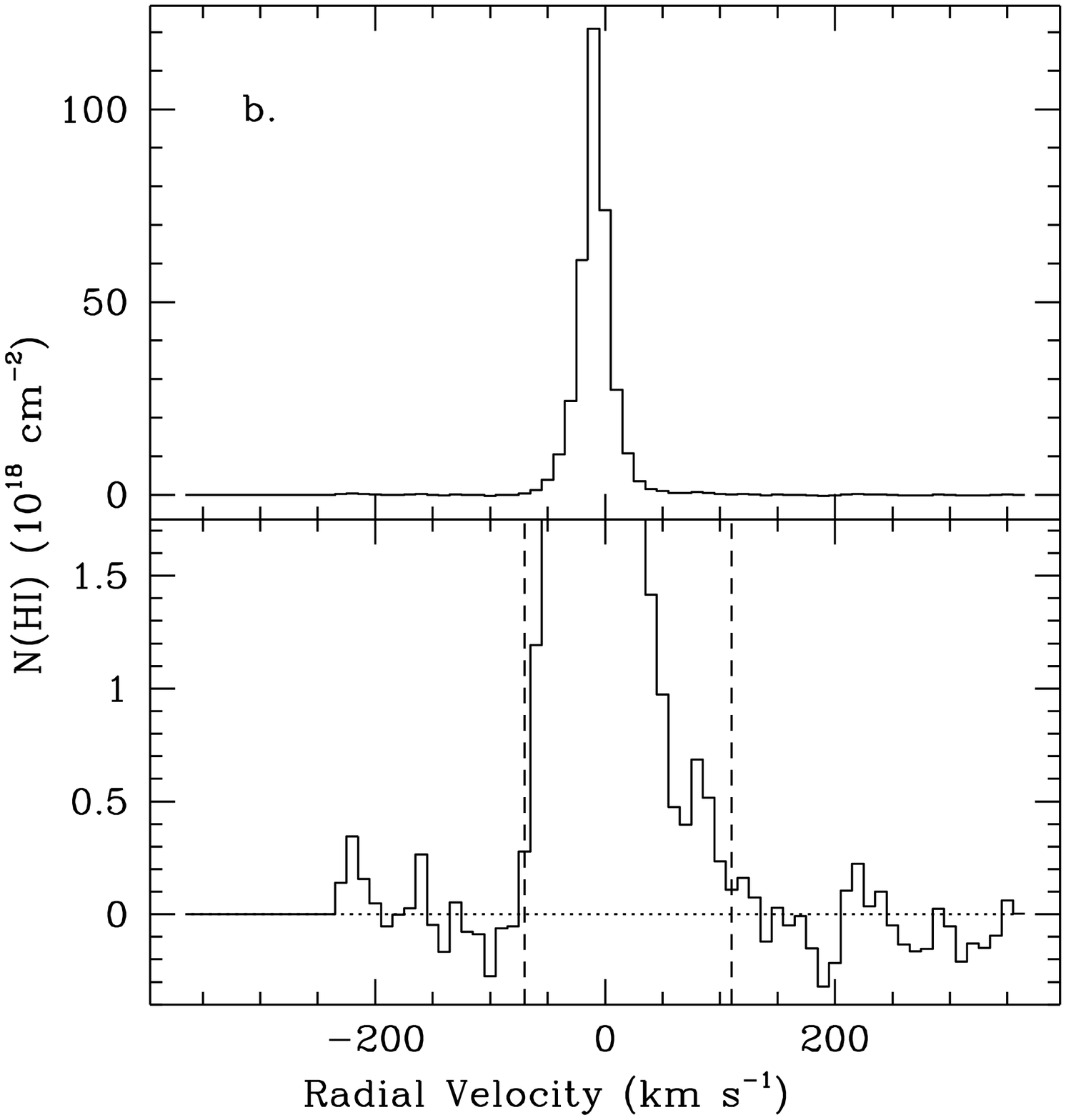}
  
 \vspace{.1in}
  
 \plottwo{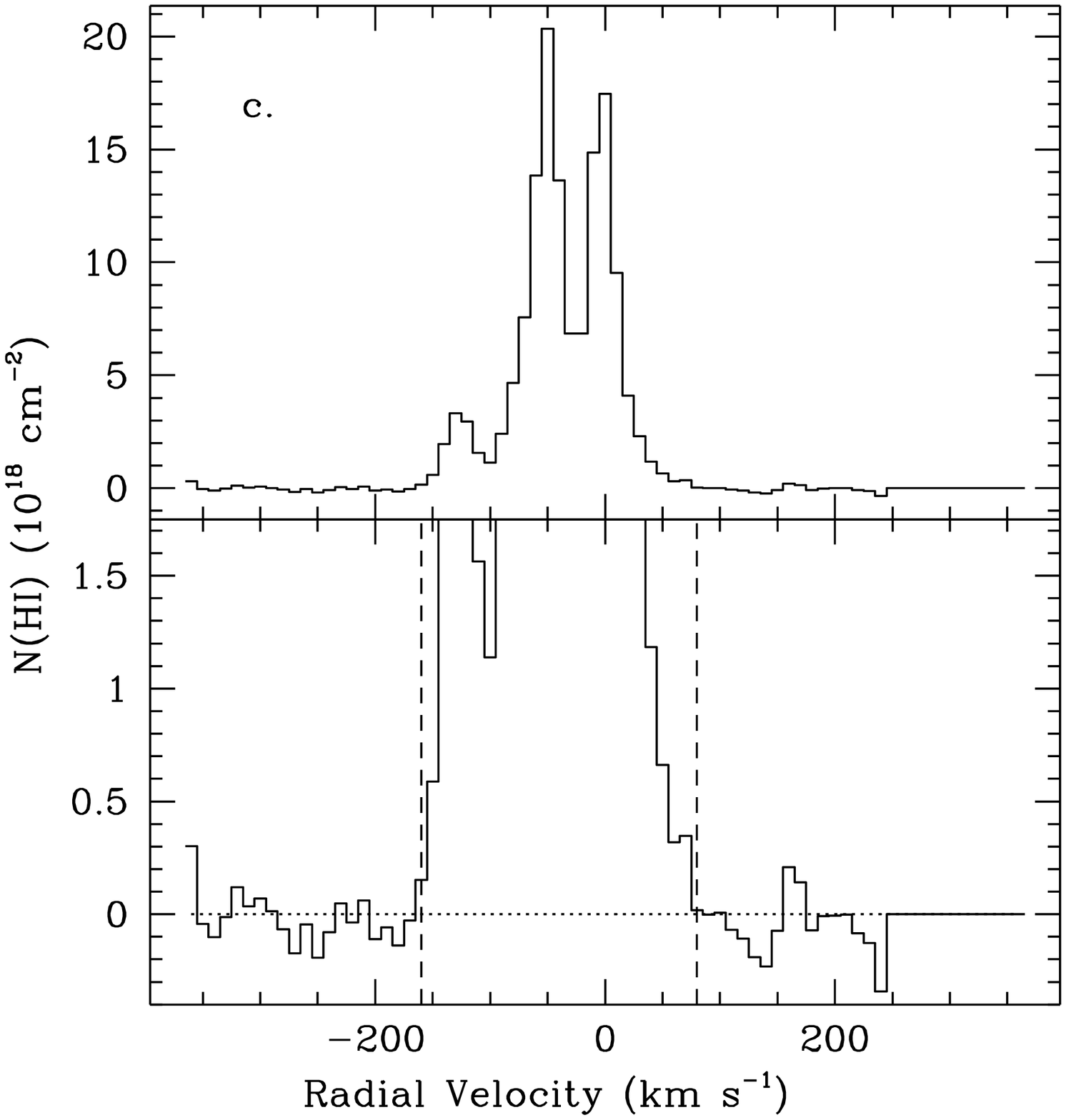}{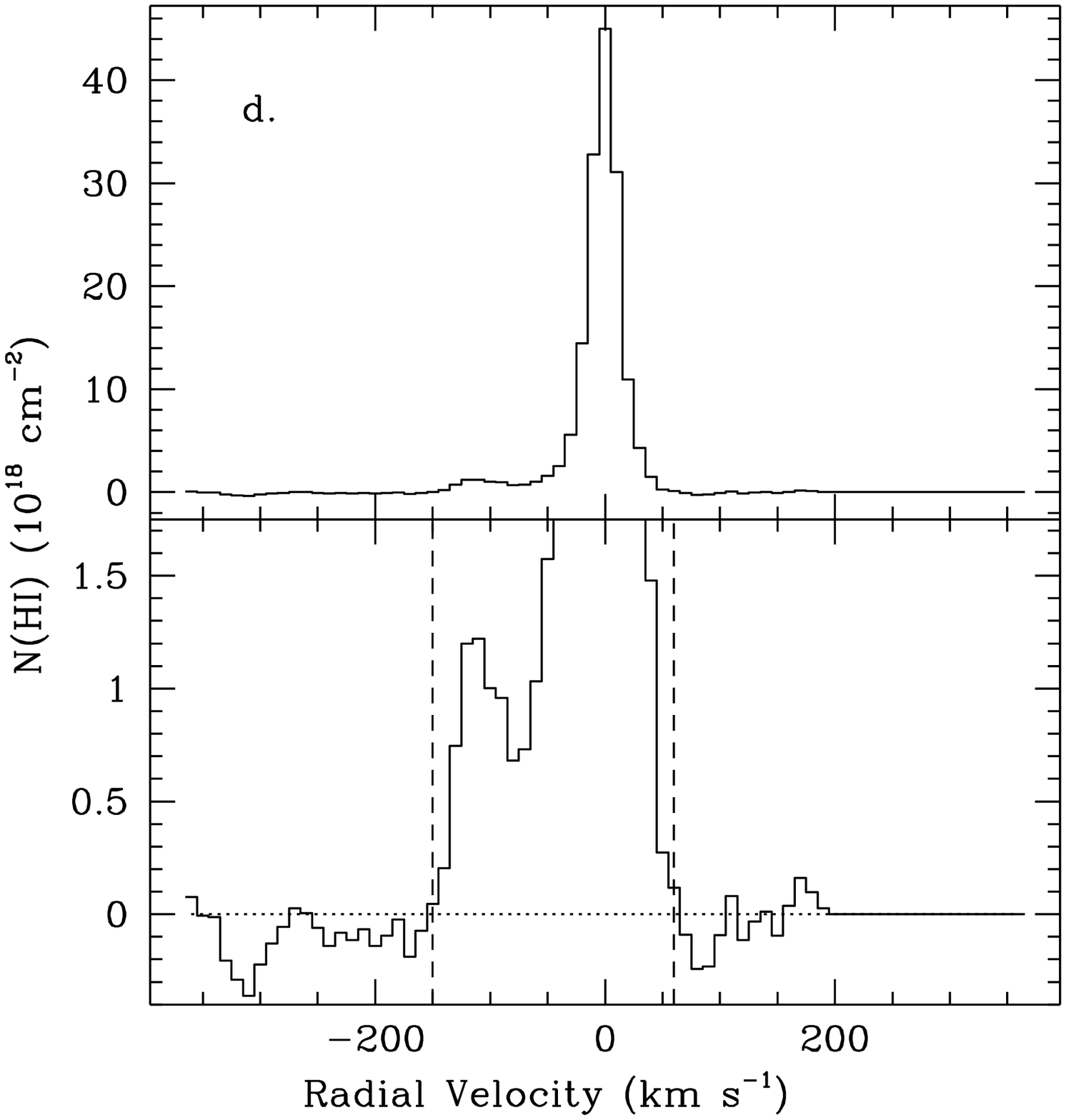}
 \caption{H\,{\sc i} 21 cm profiles from Stark et al.\  
 for the four starburst sight lines of $(a)$ IRAS 08339+6517, $(b)$ Mrk 1267, 
 $(c)$ Mrk 66, and $(d)$ Mrk 496.
 Vertical dashed lines indicate the velocity range
 within which we assume that the H\,{\sc i} detections are secure.}
 \end{figure}
%
% FIG 2
 \begin{figure}
 \plotone{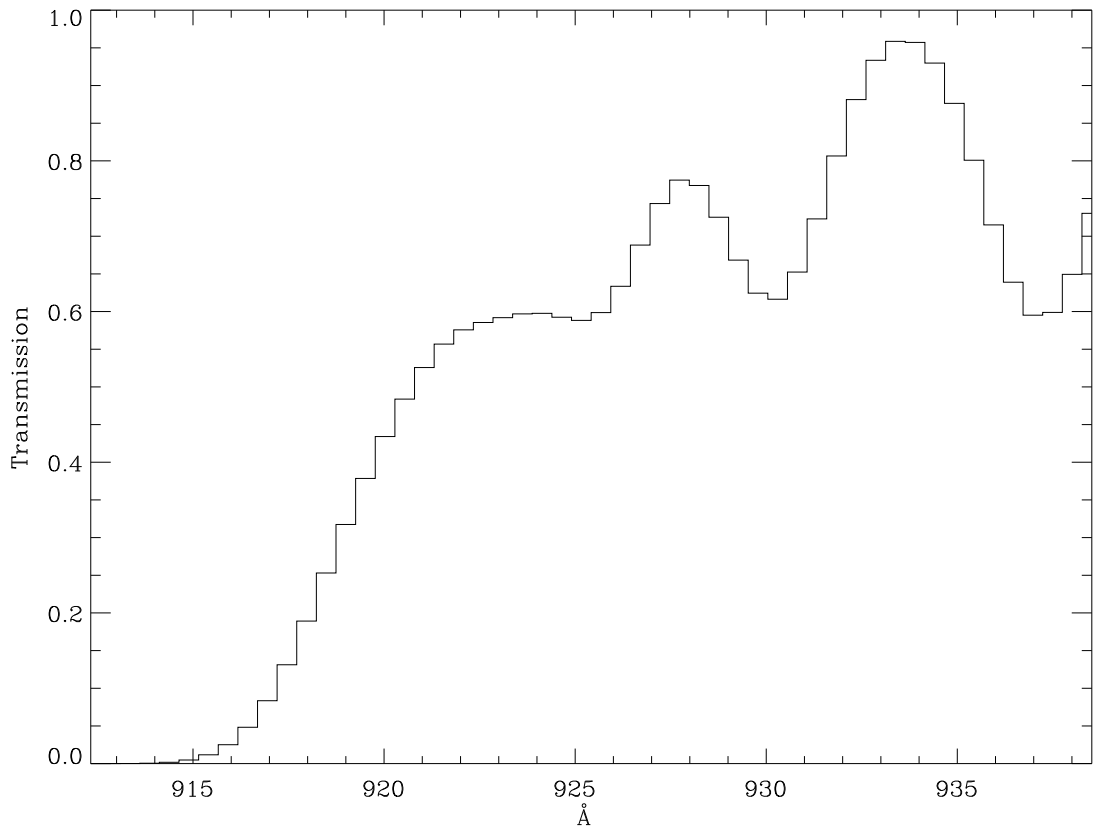}
 \caption{Gas phase interstellar medium transmission
 curve calculated for the sight line toward IRAS 08339+6517.
 Theoretical transmission has been convolved with a 3~\AA\
 FWHM Gaussian and rebinned at 0.51 \AA.
 Only H~I and associated metal lines are included.}
 \end{figure}
%
%
%%%%%%%%%%%%%%%%%%%%%%%%%%%%%%%%%%%%%%%%%%%%%%%%%%%%%%%%%%%%%%%%%%

\end{document}